\pdfoutput=1

\documentclass[11pt]{article}

\usepackage{amsmath}
\usepackage{float}
\usepackage{listings}
\usepackage{xcolor}
\usepackage{hyperref}
\usepackage[preprint]{acl}

\usepackage{times}
\usepackage{latexsym}

\usepackage[T1]{fontenc}

\usepackage[utf8]{inputenc}

\usepackage{microtype}

\usepackage{inconsolata}

\usepackage{graphicx}

\title{\includegraphics[width=0.04\textwidth]{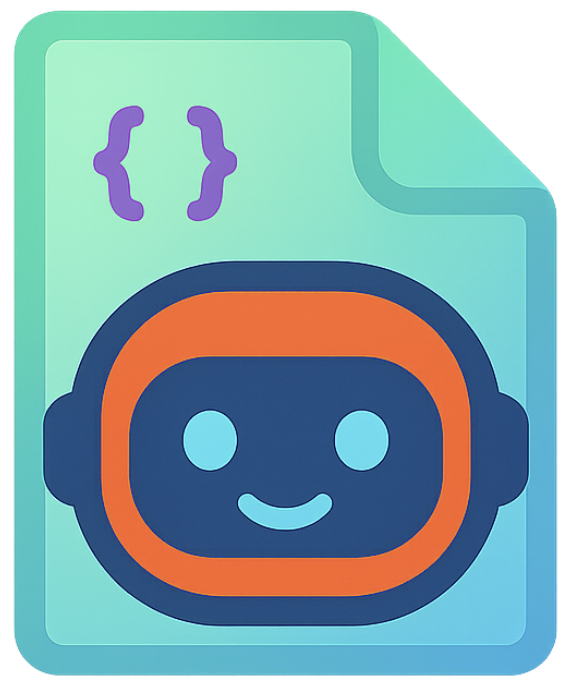} DocAgent: A Multi-Agent System for Automated Code Documentation Generation}

\author{
  Dayu Yang\thanks{Corresponding Author.} \quad
  Antoine Simoulin\thanks{Equal contribution.} \quad
  Xin Qian\footnotemark[2] \quad
  Xiaoyi Liu\footnotemark[2] \quad
  Yuwei Cao\footnotemark[2] \quad
  Zhaopu Teng\footnotemark[2] \quad
  Grey Yang \\
  Meta AI \\
  \texttt{\{dayuyang,antoinesimoulin,xinqian,xiaoyiliu,yuweicao,zhaoputeng,glyang\}@meta.com}
}

\begin{document}
\maketitle
\begin{abstract}
High-quality code documentation is crucial for software development especially in the era of AI. However, generating it automatically using Large Language Models (LLMs) remains challenging, as existing approaches often produce incomplete, unhelpful, or factually incorrect outputs.  We introduce DocAgent, a novel multi-agent collaborative system using topological code processing for incremental context building. Specialized agents (Reader, Searcher, Writer, Verifier, Orchestrator) then collaboratively generate documentation. We also propose a multi-faceted evaluation framework assessing Completeness, Helpfulness, and Truthfulness. Comprehensive experiments show DocAgent significantly outperforms baselines consistently. Our ablation study confirms the vital role of the topological processing order. DocAgent offers a robust approach for reliable code documentation generation in complex and proprietary repositories. Our code\footnote{\url{https://github.com/facebookresearch/DocAgent}} and video\footnote{\url{https://youtu.be/e9IjObGe9_I}} are publicly available.

\end{abstract}

\section{Introduction}

High-quality code documentation is essential for effective software development~\citep{de2005study, garousi2015usage, chen2009empirical}, and has become increasingly important as AI models depend on accurate docstrings\footnote{We use "code documentation" and "docstring" interchangeably throughout the paper.} for code comprehension tasks~\citep{zhou2022docprompting, yang2024less, anthropic2025modelcontext}. However, creating and maintaining documentation is labor-intensive and prone to errors~\citep{mcburney2017towards, parnas2010precise}. Even top-starred open-source repositories on GitHub often exhibit low docstring coverage and quality,\footnote{See Appendix~\ref{appendix:scarcity} for more details.} leading to documentation that frequently lags behind code changes~\citep{aghajani2019software, robillard2009makes, uddin2021automatic}.

While LLM-based solutions—such as Fill-in-the-Middle (FIM) predictors~\cite{roziere2023code, githubcopilot2024} and chat agents~\cite{metaai, openai_chatgpt}—offer automation, extensive studies~\citep{dvivedi2024comparative, zhang2024hallucinations, zan2022private, zheng2024multicodebench}, along with our empirical analyses (\S\ref{section:exp}), reveal three recurring limitations. First, these approaches often omit essential information (e.g., parameter or return-value descriptions). Second, they typically offer minimal context or rationale, limiting the usefulness of the generated documentation. Third, they sometimes hallucinate non-existent components, especially in large or proprietary repositories, undermining factual correctness~\citep{zan2022private, ma2024understand, abedu2024llm}.

\begin{figure*}
    \centering
    \includegraphics[width=1\linewidth]{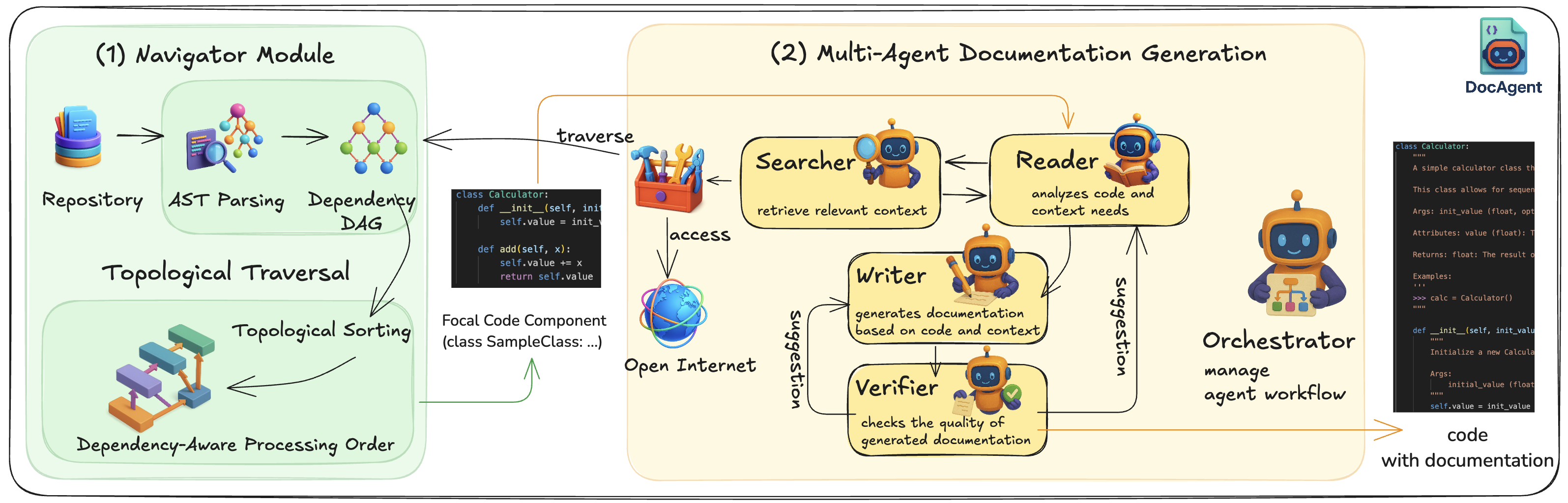}
    \caption{Architecture of DocAgent: (1) The Navigator Module uses AST parsing for a Dependency DAG and topological traversal. (2) The Multi-Agent framework uses specialized agents (Reader, Searcher, Writer, Verifier) with tools for context-aware documentation generation.}
    \label{fig:structure}
    \vspace{-4mm}
\end{figure*}

We identify three primary challenges that drive these shortcomings. \textbf{(1)~Context Identification and Retrieval:} Large, complex repositories make it non-trivial to pinpoint which files, dependencies, or external references are genuinely relevant for a given component. \textbf{(2)Navigating Complex Dependencies:} Codebases often exhibit dependency chains that exceed typical LLM context limits, requiring strategic context management. \textbf{(3)Robust and Scalable Evaluation:} Existing evaluation metrics like BLEU or ROUGE\citep{roy2021reassessing, guelman2024using} incompletely capture the multifaceted goals of documentation, while human evaluation, though more reliable, is expensive and subjective\citep{luo2024repoagent}.

To tackle these challenges, we introduce \textbf{DocAgent}, a multi-agent system that processes code in a topologically sorted order and leverages specialized agents (\textbf{Reader}, \textbf{Searcher}, \textbf{Writer}, \textbf{Verifier}, \textbf{Orchestrator}) to collaboratively generate documentation. This mimics human workflows and manages context effectively. We also propose an automatic and robust multi-faceted evaluation framework assessing \textbf{Completeness}, \textbf{Helpfulness}, and \textbf{Truthfulness} via deterministic checks and LLM-as-judge. Our main contributions are:
1) DocAgent, A multi-agent, topologically structured system for context-aware documentation generation.
2) A robust evaluation framework measuring completeness, helpfulness, and factual consistency of code documentation.
3) Comprehensive experiments on diverse repositories show DocAgent consistently outperforms state-of-the-art baselines.

\section{Methodology}
\label{sec:methodology}

DocAgent operates in two stages to handle complex dependencies and ensure context relevance. First, the \textit{Navigator} determines an optimal, dependency-aware processing order (\S\ref{sec:navigator}). Second, a \textit{Multi-Agent System} incrementally generates documentation, leveraging specialized agents for code analysis, information retrieval, drafting, and verification  (\S\ref{sec:multi_agent}). Figure~\ref{fig:structure} illustrates this architecture.

\subsection{Navigator: Dependency-Aware Order}
\label{sec:navigator}

Generating accurate documentation often requires understanding its dependencies. However, naively including the full context of all direct and transitive dependencies can easily exceed context window limit especially in large, complex repositories. To address this, the \textit{Navigator} module establishes a processing order that ensures components are documented only after their dependencies have been processed, thereby enabling incremental context building.


\noindent\textbf{Dependency Graph Construction}.
\label{sec:dep_graph}
DocAgent first performs static analysis on the entire target repository. It parses the Abstract Syntax Trees (ASTs) of source files to identify code components (functions, methods, classes) and their interdependencies. These dependencies include function/method calls, class inheritance, attribute access, and module imports. These components and relationships are used to construct a directed graph where nodes represent code components and a directed edge from A to B signifies that A depends on B (A $\rightarrow$ B). To enable topological sorting, cycles within the graph are detected using Tarjan's algorithm~\cite{tarjan1972depth} and condensed into a single super node. This results in a Directed Acyclic Graph (DAG) representing the repository's dependency structure.

The process begins with static analysis of the entire target repository. Abstract Syntax Trees (ASTs) are parsed for all source files to identify core code components (e.g., functions, methods, classes) and their interdependencies. These dependencies encompass function/method calls, class inheritance relationships, attribute accesses, and module imports. Based on this analysis, a directed graph is constructed where nodes represent code components and a directed edge from component A to component B (A $\rightarrow$ B) signifies that A depends on B (i.e., B must be understood to fully understand A)\footnote{Cycles within the graph are detected using Tarjan's algorithm~\citep{tarjan1972depth} and condensed into a single node.}.

\noindent\textbf{Topological Traversal for Hierarchical Generation}.
\label{sec:topo_traversal}
Using the DAG, the Navigator performs a topological sort to determine the documentation generation order. The traversal adheres to the "Dependencies First" principles: A component is processed only after all components it directly depends on have been documented\footnote{Methods are documented before their enclosing class.}. This topological ordering ensures that, by the time the multi-agent system generates documentation for a given component, all of its dependencies have already been described. Therefore, each code documentation only needs the information of its one-hop dependencies, eliminating the need to pull in an ever-growing chain of background information.

\subsection{Multi-Agent Documentation Generation}
\label{sec:multi_agent}
Following Navigator's order, the multi-agent system generates documentation for each component using four specialized agents coordinated by an Orchestrator. Input is the focal component's source code including newly generated documentation.

\begin{figure}
    \centering
    \includegraphics[width=0.99\linewidth]{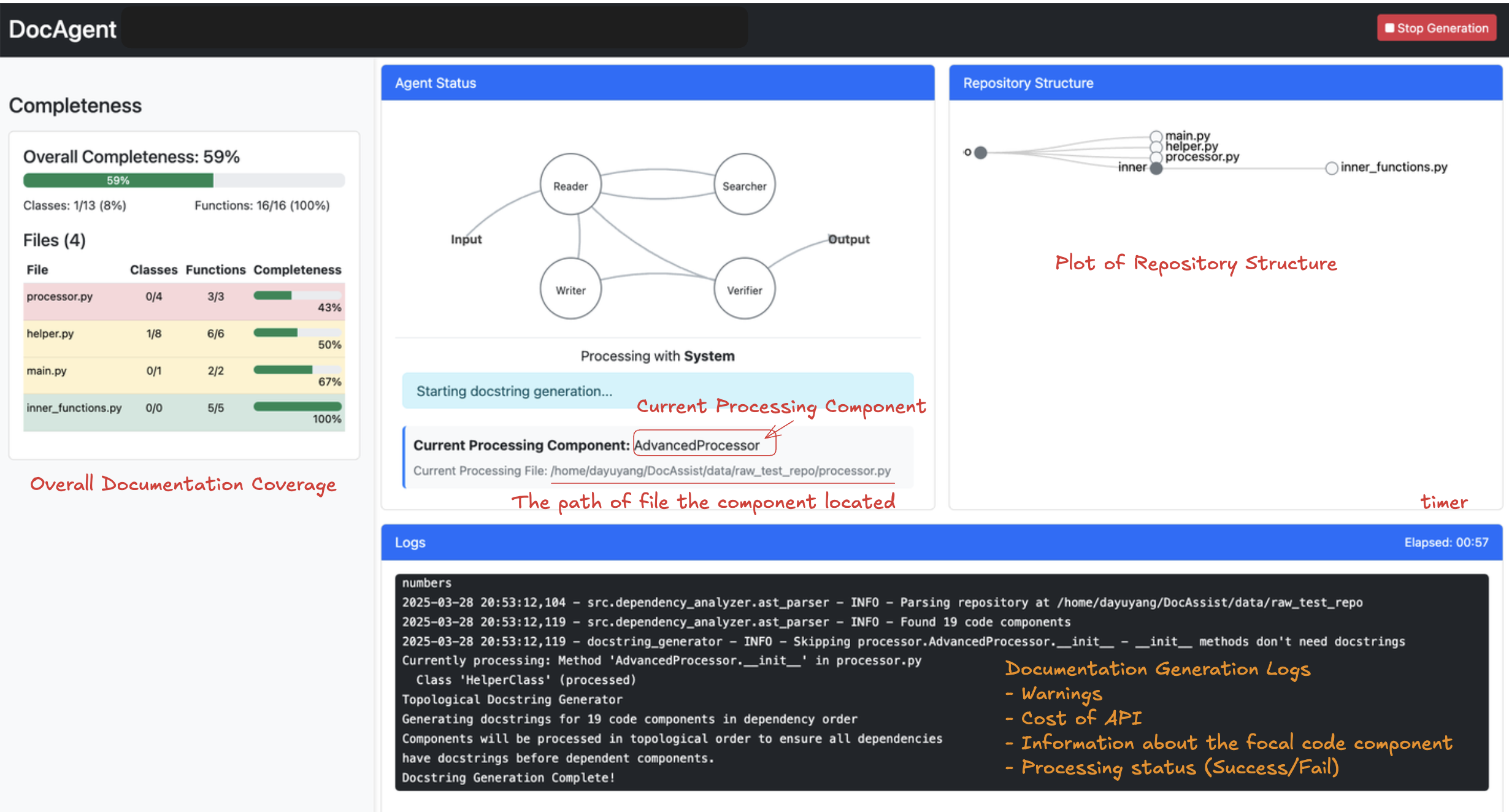}
    \caption{Screenshot of DocAgent live code documentation generation page.}
    \label{fig:sys_gen}
    \vspace{-4mm}
\end{figure}

\textbf{Reader}.
\label{sec:reader}
The Reader agent initiates the process by analyzing the focal component's code. Its primary goal is to determine the information required to generate a comprehensive and helpful code documentation. It assesses the component's complexity, visibility (public/private), and implementation details to decide:
\textit{If additional context is needed:} Simple, self-contained components might not require external information.
\textit{What context is needed:} This involves identifying specific internal dependencies (functions/classes it uses), usage contexts (where the component is called, revealing its purpose), or external concepts (algorithms, libraries, domain knowledge) referenced implicitly or explicitly.

The agent outputs structured XML requests for two types of information requests (1) internal information about related code components, and (2) external knowledge for specialized algorithms or techniques.

The internal information request consists with the dependency and the reference. Dependency means the focal component calls other components defined in the repository, where reader will determinate if a dependent is needed or not to provide necessary context information.

Reference means the focal component is called in somewhere in the code repository, showing how it can be used in the real-world application and therefore reveal the purpose of the focal code component. This is particularly important for public functions or APIs exposed to the users of the repository.

External requests target information not directly present or inferable from the codebase itself, such as domain-specific knowledge or third-party library functionalities (see Appendix~\ref{appendix:why_external}).

\textbf{Searcher}.
\label{sec:searcher}
The Searcher agent is responsible for fulfilling the Reader's information requests using specialized tools:
\textit{Internal Code Analysis Tool:} This tool leverages static analysis capabilities to navigate the codebase. It can retrieve the source code and existing documentation of specified internal components, identify call sites for the focal component, trace dependencies using the pre-computed graph or on-the-fly analysis, and extract relevant structural information (e.g., class hierarchies, method signatures).
\textit{External Knowledge Retrieval Tool:} This tool interfaces with external knowledge sources via a generic retrieval API . It formulates queries based on the Reader's requests for external concepts and processes the results to extract pertinent explanations, definitions, or descriptions.

The Searcher consolidates the retrieved internal code information and external knowledge into a structured format, which serves as the context for the subsequent agents.

Like two human agents collaborate on a project and talk with each other, after Searcher send the retrieved information back to the reader, reader read the updated context and the focal code component, and see if the context is adequate for generating the documenation. If reader still feel the retrieved context is still not adequate, reader can further send information request to the searcher. So the information request, and new information can be sent back and forth between reader and searcher, until adequate information is retrieved.

\textbf{Writer}.
\label{sec:writer}
The Writer agent receives the focal component's code and the structured context compiled by the Searcher. Its task is to generate the code documentation. The generation process is guided by prompts that specify the desired structure and content based on the component type:
\textit{Functions/Methods:} Typically require a summary, extended description, parameter descriptions (Args), return value description (Returns), raised exceptions (Raises), and potentially usage examples (especially for public-facing components).
\textit{Classes:} Typically require a summary, extended description, initialization examples, constructor parameter descriptions (Args), and public attribute descriptions (Attributes).

The Writer synthesizes information from both the code and the provided context to produce a draft code documentation adhering to these requirements.

\textbf{Verifier}.
\label{sec:verifier}
The Verifier take the context, code component, and generated code documentation from the writer as inputs, evaluates the quality of code documentation against predefined criteria: information value, detail level, and completeness. Upon evaluation, the Verifier either approves the documentation or provides specific improvement suggestions through structured feedback. 

Verifier can talk to writer if the issue can be address without additional context information, for example: format issue, which can be easily address by asking writer to rewrite. 

If the issue is relevant to lack of information, and additional context is needed, veirfier can also provide suggestion to reader, and additional information will be gathered through another Reader-Searcher cycle.

\textbf{Orchestrator}.
\label{sec:orchestrator}
An Orchestrator manages the agent workflow through an iterative process. The cycle begins with the Reader analyzing the focal component and requesting necessary context. The Searcher gathers this information, after which the Writer generates a docstring. The Verifier then evaluates the docstring quality, either approving it or returning it for revision. This process continues until a satisfactory code documentaion is generated or a maximum iteration limit is reached.

\textit{Adaptive Context Management:} To handle potentially large contexts retrieved by the Searcher, especially for complex components, the Orchestrator implements an adaptive context truncation mechanism. It monitors the total token count of the context provided to the Writer. If the context exceeds a configurable threshold (based on the underlying LLM's limits), the Orchestrator applies a targeted truncation strategy. It identifies the largest sections within the structured context (e.g., external knowledge snippets, specific dependency details) and selectively removes content from the end of these sections to reduce the token count while preserving the overall structure. This ensures that the context remains within operational limits, balancing contextual richness with model constraints.

\begin{figure}
    \centering
    \includegraphics[width=1\linewidth]{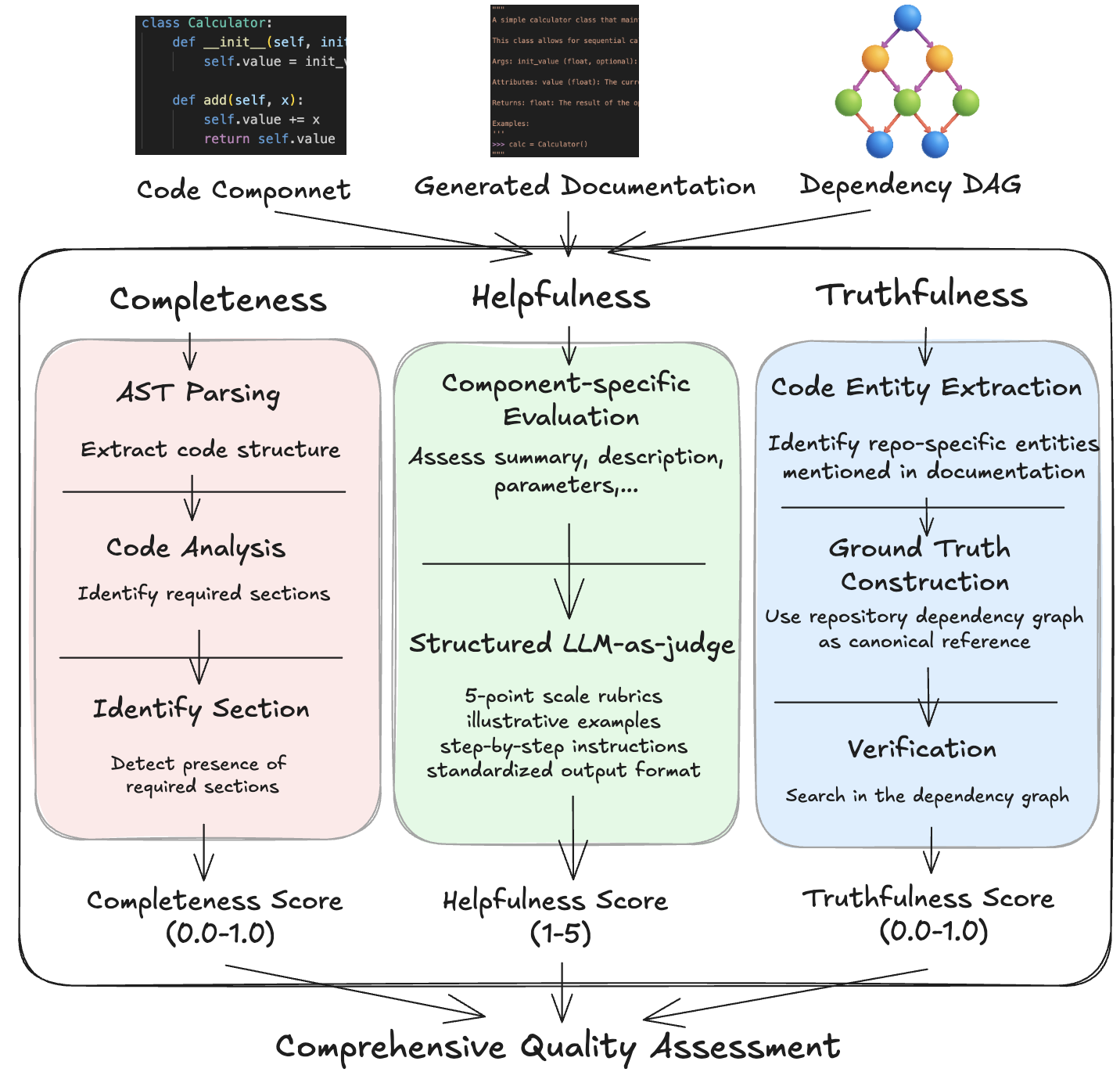}
    \caption{Multi-facet Evaluation Framework of code documentation, assessing quality along three dimensions: (1) Completeness measures structural adherence to documentation conventions; (2) Helpfulness evaluates practical utility; and (3) Truthfulness verifies factual accuracy.}
    \label{fig:evaluation}
    \vspace{-5mm}
\end{figure}

\section{Evaluation Framework}
\label{sec:evaluation}

Evaluating the quality of automatically generated code documentation is challenging. Traditional metrics commonly used in natural language generation, such as BLEU or ROUGE cannot be used because of lack of gold references~\cite{roy2021reassessing, guelman2024using}. Simple heuristics like documentation length are insufficient indicators of actual utility. While human evaluation provides the most accurate assessment~\cite{luo2024repoagent}, it is inherently subjective, expensive, and difficult to scale, rendering it impractical for large-scale experiments or continuous integration scenarios.

To overcome these limitations, we propose a comprehensive and scalable evaluation framework designed to systematically assess documentation quality along three crucial dimensions: \textit{Completeness}, \textit{Helpfulness}, and \textit{Truthfulness}. This multi-faceted approach combines deterministic structural checks, LLM-based qualitative assessments, and fact-checking against the codebase itself, providing a holistic view of the generated documentation's value. Our methodology is informed by established software engineering best practices for documentation and addresses the specific shortcomings observed in existing LLM-based generation systems.

\begin{figure}
    \centering
    \includegraphics[width=0.99\linewidth]{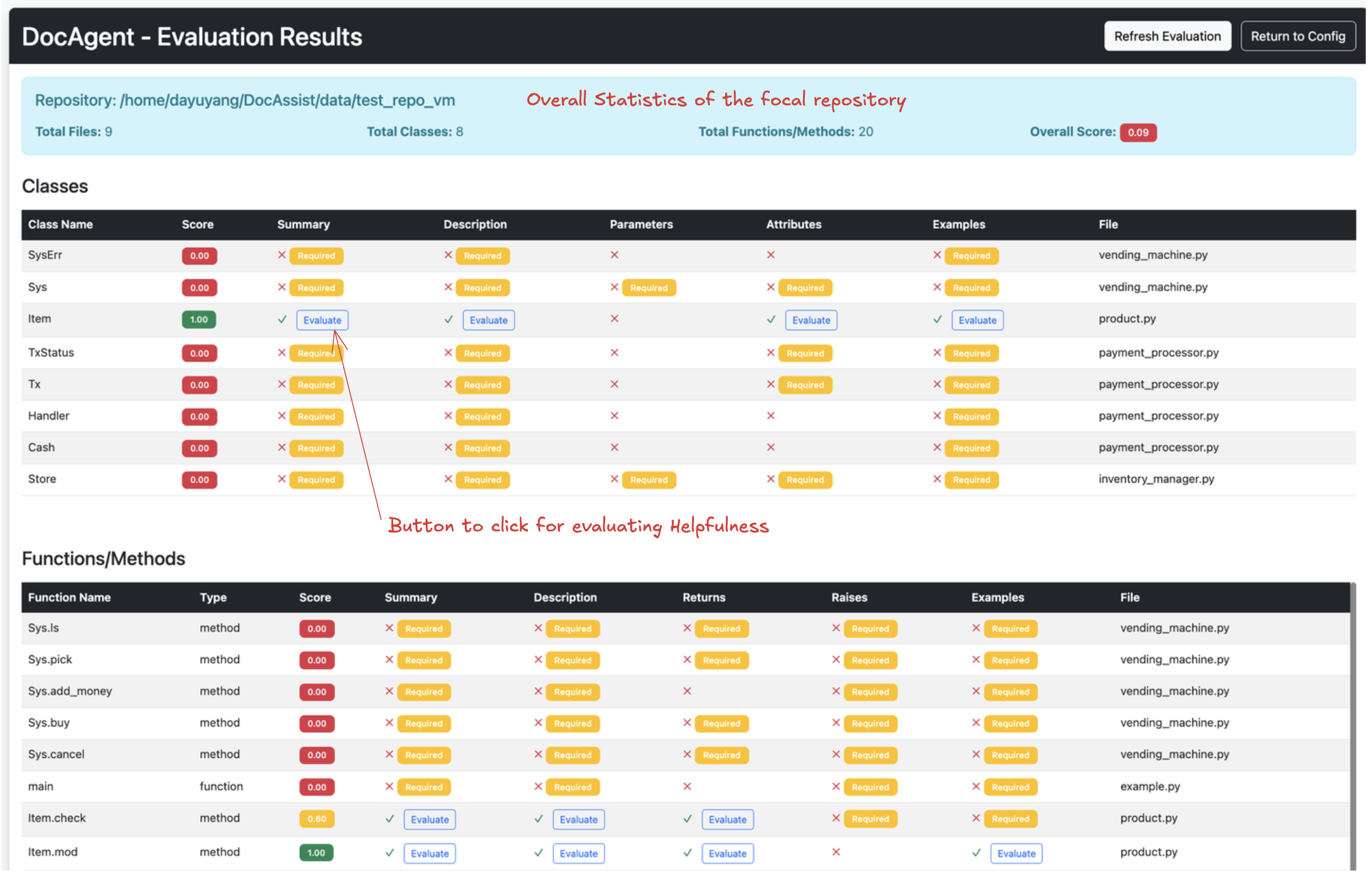}
    \caption{Screenshot of DocAgent Live Evaluation Framework}
    \label{fig:sys_eval}
        \vspace{-4mm}
\end{figure}

\subsection{Completeness}
\label{sec:eval_completeness}

Completeness measures the extent to which the generated documentation adheres to standard structural conventions and includes essential components expected for a given code element (e.g., function, class). High-quality code documentation typically includes not only a summary but also descriptions of parameters, return values, raised exceptions, and potentially usage examples, dynamically depending on the element's signature, body and visibility.

To quantify completeness, we employ an automated checker based on Abstract Syntax Tree (AST) analysis and regular expressions. The process involves:
\textbf{AST Parsing:} Identifying code components (classes, functions, methods) and extracting their generated docstrings.
\textbf{Code Analysis:} Analyzing the code signature and body (e.g., presence of parameters, return statements, \texttt{raise} statements) and visibility (public/private) to determine the \textit{required} documentation sections dynamically. For instance, a function without parameters does not require an "Args" section, while a public class method might benefit more from an "Example" section than a private helper function.
\textbf{Section Identification:} Detecting the presence of standard sections (e.g., Summary, Description, Args, Returns, Raises, Examples, Attributes for classes) within the docstring using predefined patterns and structural cues.
\textbf{Scoring:} Calculating a completeness score for each docstring as the proportion of required sections that are present. This yields a normalized score between 0.0 and 1.0.

This deterministic approach provides an objective measure of structural adherence, indicating whether the documentation meets basic formal requirements.

\subsection{Helpfulness}
\label{sec:eval_helpfulness}

Helpfulness assesses the semantic quality and practical utility of the documentation content. A helpful docstring goes beyond merely restating code elements; it elucidates the \textit{purpose}, \textit{usage context}, \textit{design rationale}, and potential \textit{constraints} of the code. Key aspects include:
\textbf{Clarity and Conciseness:} Is the summary informative yet brief?
\textbf{Descriptive Depth:} Does the extended description provide sufficient context, explain the 'why' behind the code, or mention relevant scenarios or edge cases?
\textbf{Parameter/Attribute Utility:} Are descriptions for inputs and attributes meaningful, specifying expected types, value ranges, or constraints, rather than just echoing names?
\textbf{Guidance:} Does the documentation effectively guide a developer on \textit{when} and \textit{how} to use the component?

Assessing these qualitative aspects automatically is challenging. Inspired by recent work on evaluating complex generation tasks~\cite{wang2024autosurvey, zhuge2024agent},
we utilize an LLM-as-judge approach, carefully structured to enhance robustness and consistency. To mitigate potential biases and variability associated with LLM judgments, we implement a sophisticated framework:
\textbf{Component-Specific Evaluation:} We decompose the evaluation by assessing distinct parts of the docstring separately (e.g., summary, main description, parameter descriptions) using tailored prompts for each.
\textbf{Structured Prompt Engineering:} Each prompt includes:
1) \textit{Explicit Scoring Rubrics:} Detailed criteria for a 5-point Likert scale (1=Poor to 5=Excellent), defining expectations for each score level regarding clarity, depth, and utility.
2) \textit{Illustrative Examples:} Concrete examples of good and bad documentation snippets corresponding to different score levels, grounding the evaluation criteria.
3) \textit{Step-by-Step Instructions:} Guiding the LLM to analyze the code, compare the docstring against the rubric, consider the code's context, and justify its rating.
4) \textit{Standardized Output Format:} Requiring the LLM to provide structured output, including detailed reasoning, specific suggestions for improvement (if applicable), and the final numerical score. This facilitates analysis and consistency checking.

This structured approach allows for scalable assessment of semantic quality, moving beyond surface-level checks to gauge the documentation's actual value to a developer.

\subsection{Truthfulness}
\label{sec:eval_truthfulness}

A critical dimension of documentation quality is its factual accuracy, or \textit{Truthfulness}. Documentation, especially when generated by LLMs unfamiliar with a specific private codebase, can suffer from "hallucinations"—confidently referencing non-existent methods, parameters, or classes, or misrepresenting relationships between components. Such inaccuracies severely undermine trust and can mislead developers.

We evaluate Truthfulness by verifying whether entities mentioned in the generated documentation actually exist within the target repository and are referenced correctly. Our pipeline comprises three stages:
\textbf{Code Entity Extraction:} An LLM is prompted to identify mentions of repository-specific code components (classes, functions, methods, attributes) within the generated docstring. The prompt specifically instructs the model to distinguish these from standard language keywords, built-in types (e.g., \texttt{list}, \texttt{dict}), and common external library components, focusing on internal references.
\textbf{Ground Truth Construction:} We leverage the dependency graph constructed by the Navigator module~\ref{sec:dep_graph}. 
    This graph serves as the ground truth, containing a canonical representation of all code components and their locations within the repository.
\textbf{Verification:} Each extracted entity mention is cross-referenced against the dependency graph. 

We quantify Truthfulness using the \textbf{Existence Ratio}: the proportion of unique repository-specific entities mentioned in the documentation that correspond to actual entities in the codebase.$
\text{Existence Ratio} = \frac{|\text{Verified Entities}|}{|\text{Extracted Entities}|}
$

A high ratio indicates that the documentation is well-grounded in the actual code structure, minimizing the risk of hallucinated references.

Together, these three dimensions—Completeness, Helpfulness, and Truthfulness—provide a robust and nuanced framework for evaluating automatic code documentation systems, enabling quantitative comparisons and deeper insights into their strengths and weaknesses.

\section{Experiment}
\label{section:exp}

\subsection{Baselines}
We compare DocAgent against two representative baseline systems commonly used for code documentation generation:
\noindent\textbf{FIM} (Fill-in-the-middle): Simulates inline code completion tools that predict documentation based on surrounding code. We use CodeLlama-13B~\cite{roziere2023code}, an open model trained with FIM tasks~\cite{bavarian2022efficient}. Abbreviated as \textbf{FIM-CL}.
\noindent\textbf{Chat}: Represents generating documentation by providing the code snippet directly to a chat-based LLM. We test two leading models: GPT-4o mini~\footnote{2024-07-18 version}\cite{openai_chatgpt} and CodeLlama-34B-instruct\cite{roziere2023code}. Abbreviated as \textbf{Chat-GPT} and \textbf{Chat-CL}, respectively.

\subsection{Experiment Setup}

\noindent\textbf{Data.} We select a representative subset of Python repositories to ensure diversity in size, complexity, and domain. The dataset comprises modules, functions, methods, and classes with varying degrees of dependency density (details in Appendix~\ref{appendix:data}).

\noindent\textbf{Systems.} We evaluate two variants of our proposed system, differing only in the backbone LLM used by the agents: \textbf{DA-GPT}: DocAgent utilizing GPT-4o mini. \textbf{DA-CL}: DocAgent utilizing CodeLlama-34B-instruct\footnote{The choice of backbone LLM is orthogonal to the DocAgent framework itself. We use GPT-4o-2024-08-06 universally for running evaluation for more robust results.}.

\noindent\textbf{Statistical Significance.} All claims of statistical significance are based on paired t-tests with a significance threshold of $p < 0.05$\footnote{Due to space limitations, we are unable to include the full prompts and detailed experimental setup in the paper. However, all configurations are available in our project’s public release repository.}

\subsection{Experiment Results}
\label{section:experiment_results}

We evaluate the systems using the framework proposed in Section~\ref{sec:evaluation}, focusing on Completeness, Helpfulness, and Truthfulness.

\subsubsection{Completeness}
\label{sec:results_completeness}

\begin{table}[ht]
\centering
\resizebox{\columnwidth}{!}{%
\begin{tabular}{@{}lcccc@{}} 
\hline
System & Overall & Function & Method & Class \\ \hline
\textbf{DA-GPT} & 0.934\textsuperscript{\dag} & 0.945\textsuperscript{\dag} & 0.935\textsuperscript{\dag} & \textbf{0.914}\textsuperscript{\dag} \\
\textbf{DA-CL} & \textbf{0.953}\textsuperscript{\dag\ddag} & \textbf{0.985}\textsuperscript{\dag\ddag} & \textbf{0.982}\textsuperscript{\dag\ddag} & 0.816\textsuperscript{\dag\ddag} \\
Chat-GPT & 0.815 & 0.828 & 0.823 & 0.773 \\
Chat-CL & 0.724 & 0.726 & 0.744 & 0.667 \\
FIM-CL & 0.314 & 0.291 & 0.345 & 0.277 \\\hline
\end{tabular}%
}
\caption{Average Completeness Scores. \textsuperscript{\dag}: Significantly better than corresponding Chat baseline. \textsuperscript{\ddag}: Significantly better than FIM baseline.}
\label{tab:completeness_results}
\vspace{-2mm}
\end{table}

As shown in Table~\ref{tab:completeness_results}, both DocAgent variants significantly outperform their respective Chat counterparts. DocAgent (CodeLlama-34B) achieves an overall score of 0.953, representing a substantial improvement of 0.229 points over Chat. Similarly, DocAgent (GPT-4o mini) scores 0.934 overall, significantly higher than Chat at 0.815. These improvements are statistically significant across all component types. FIM performs poorly, achieving an overall completeness score of only 0.314. 
This highlights the effectiveness of DocAgent's structured, context-aware generation process compared to simply prompting an LLM with the code in isolation.

\subsubsection{Helpfulness}

As shown in Table~\ref{tab:helpfulness_results}, DocAgent (GPT-4o mini) achieves the highest overall helpfulness score, significantly outperforming the corresponding Chat baseline. demonstrating its ability to generate clearer and more informative content by leveraging retrieved context.

\begin{table}[ht]
\centering
\resizebox{\columnwidth}{!}{%
\begin{tabular}{@{}lcccc@{}} 
\hline
System & Overall & Summary & Description & Parameters \\ \hline
\textbf{DA-GPT} & \textbf{3.88}\textsuperscript{\dag} & \textbf{4.32}\textsuperscript{\dag} & \textbf{3.60}\textsuperscript{\dag} & \textbf{2.71} \\
\textbf{DA-CL} & 2.35\textsuperscript{\ddag} & 2.36\textsuperscript{\dag\ddag} & 2.43\textsuperscript{\ddag} & 2.00 \\
Chat-GPT & 2.95 & 3.56 & 2.42 & 2.20 \\
Chat-CL & 2.16 & 2.04 & 2.37 & 1.80 \\
FIM-CL & 1.51 & 1.30 & 2.45 & 1.50 \\ \hline
\end{tabular}%
}
\caption{Average Helpfulness Scores. \textsuperscript{\dag}: Significantly better than corresponding Chat. \textsuperscript{\ddag}: Significantly better than FIM.}
\label{tab:helpfulness_results}
\vspace{-2mm}
\end{table}

DocAgent (CodeLlama-34B) also shows an improvement over its Chat counterpart, producing significantly more helpful summaries. Furthermore, DocAgent (CodeLlama-34B) also significantly outperforms FIM. Across aspects, generating helpful parameter descriptions appears most challenging. DocAgent (GPT-4o mini) achieves the highest score even here, suggesting its structured approach aids in this difficult task, although room for improvement remains.

\subsubsection{Truthfulness}
\label{sec:results_truthfulness}

The results in Table~\ref{tab:truthfulness_results} demonstrate the superior factual accuracy of documentation generated by DocAgent. DocAgent (GPT-4o mini) achieves the highest Existence Ratio at 95.74\%, indicating that the vast majority of its references to internal code components are correct. DocAgent (CodeLlama-34B) also performs strongly with a ratio of 88.17\%.

\begin{table}[ht]
\centering

\resizebox{\columnwidth}{!}{%
\begin{tabular}{@{}lccc@{}} 
\hline
System &  Verified &  Extracted & Existence Ratio (\%) \\ \hline
\textbf{DA-GPT} & 265 & 305 & \textbf{95.74\%} \\
\textbf{DA-CL} & 354 & 600 & 88.17\% \\
Chat-GPT & 366 & 347 & 61.10\% \\
Chat-CL & 366 & 488 & 68.03\% \\
FIM-CL & 338 & 131 & 45.04\% \\ \hline
\end{tabular}%
}
\caption{Truthfulness Analysis: Existence Ratio (\%). Higher is better. Extracted = extracted entities; Verifed = verified entities in \S\ref{sec:eval_truthfulness}.}
\label{tab:truthfulness_results}
\end{table}

This contrasts sharply with the baselines. The Chat approaches exhibit significantly lower truthfulness, with Chat (GPT-4o mini) at 61.10\% and Chat (CodeLlama-34B) at 68.03\%. This suggests that simply providing the code snippet to a chat model often leads to inaccurate assumptions or hallucinations about the surrounding codebase context. FIM performs worst, with an Existence Ratio of only 45.04\%, implying that nearly half of its references to repository entities might be incorrect. This low score highlights a significant risk of misleading developers when using FIM for documentation.

\subsection{Ablation Study}
\label{sec:ablation}

To isolate the contribution of the dependency-aware processing order determined by the Navigator module (\S~\ref{sec:navigator}), we conducted an ablation study. We created variants of DocAgent (DA-Rand-GPT, DA-Rand-CL) that process components in a random order\footnote{Completeness was omitted from the ablation study because it depends on the code's structure, not the Navigator's processing order.}.

\subsubsection{Impact on Helpfulness}

\begin{table}[ht]
\centering
\resizebox{\columnwidth}{!}{%
\begin{tabular}{@{}lcccc@{}} 
\hline
System & Overall & Summary & Description & Parameters \\ \hline
\textbf{DA-GPT} & \textbf{3.88}\textsuperscript{\dag} & \textbf{4.32}\textsuperscript{\dag} & \textbf{3.60} & \textbf{2.71} \\
DA-Rand-GPT & 3.44(-0.44) & 3.62(-0.70) & 3.30(-0.30) & 2.20(-0.51) \\ \hline
\textbf{DA-CL} & \textbf{2.35}\textsuperscript{\dag} & \textbf{2.36}\textsuperscript{\dag} & \textbf{2.43} & \textbf{2.00} \\
DA-Rand-CL & 2.18(-0.17) & 1.88(-0.48) & 2.42(-0.10) & 2.00(\phantom{-}0.00) \\ \hline
\end{tabular}%
}
\caption{Ablation: Average Helpfulness Scores. 
\textsuperscript{\dag} If DocAgent significantly better than its Random variant.}
\label{tab:ablation_helpfulness}
\end{table}

The results in Table~\ref{tab:ablation_helpfulness} demonstrate the benefit of the Navigator's topological sorting in improving Helpfulness. For both underlying LLMs, the full DocAgent achieved significantly higher overall helpfulness scores compared to its random-order counterpart.
With GPT-4o mini, the full DocAgent scored 3.69 overall, significantly higher than DocAgent-Random's 3.44. The improvement was particularly pronounced in summary generation.
Similarly, with CodeLlama-34B, the full DocAgent scored 2.39 overall, significantly outperforming DocAgent-Random's 2.18. Again, the summary scores showed a significant difference.

\subsubsection{Impact on Truthfulness}

We also evaluated the impact of removing the hierachical generation order on the factual accuracy (Truthfulness). Without the Navigator, the Searcher can still retrieve dependent code components. However, since the 'Dependencies First' principle is not followed, these components are less likely to have already generated documentation available for context.

\begin{table}[ht]
\centering
\resizebox{\columnwidth}{!}{%
\begin{tabular}{@{}lccc@{}} 
\hline
System & Verified & Extracted & Existence Ratio (\%) \\ \hline
\textbf{DA-GPT} & 187 & 224 & \textbf{94.64\%} \\
DA-Rand-GPT & 164(-23) & 166(-58) & 86.75(-7.89)\% \\ \hline
\textbf{DA-CL} & 190 & 343 & \textbf{87.76\%} \\
DA-Rand-CL & 188(-2) & 360(+17) & 83.06(-4.70)\% \\ \hline
\end{tabular}%
}
\caption{Ablation: Truthfulness Analysis (Existence Ratio \%). Use 50 randomly sampled code components from full data to evaluate.}
\label{tab:ablation_truthfulness}
\vspace{-4mm}
\end{table}

Table~\ref{tab:ablation_truthfulness} demonstrates that the topological sort also improves truthfulness. Both full DocAgent variants achieve higher Existence Ratios than their random-order counterparts. Existence ratio of DocAgent (GPT-4o-mini) drops from 94.64\% to 86.75\% without the sort, and the ratio of DocAgent (Codellama-34B) drops from 87.76\% to 83.06\%. 

Collectively, the ablation results confirm that the Navigator's dependency-aware topological ordering is a crucial component of DocAgent, significantly contributing to both the helpfulness and factual accuracy of the generated documentation by enabling effective incremental context management.

\section{Conclusion}
\label{sec:conclusion}

We addressed the challenge of automatically generating high-quality code documentation, a task where existing LLM-based methods often struggle with incompleteness, lack of helpfulness, and factual inaccuracies. We introduced \textbf{DocAgent}, a novel tool-integrated, multi-agent system that leverages a dependency-aware topological processing order determined by a \textbf{Navigator} module. This allows specialized agents (Reader, Searcher, Writer, Verifier, Orchestrator) to collaboratively generate documentation by incrementally building context from dependencies. We also proposed a robust and scalable evaluation framework assessing \textbf{Completeness}, \textbf{Helpfulness}, and \textbf{Truthfulness}. Our experiments on diverse Python repositories demonstrate that DocAgent significantly outperforms FIM and Chat baselines consistently, producing more complete, helpful, and factually accurate documentation. An ablation study confirmed the critical contribution of the topological processing order to both helpfulness and truthfulness. DocAgent represents a promising step towards reliable and useful automated code documentation generation for complex and proprietary software.

\section{Ethics and Limitations}
\label{sec:ethics_limitations}

DocAgent, while advancing automated code documentation, has inherent limitations and ethical considerations. Technically, processing extremely large codebases may still challenge LLM context limits despite topological sorting and context management. Relying solely on static analysis restricts understanding of dynamic behavior, and the current Python focus requires effort for adaptation to other languages.

Ethically, the primary concern is factual accuracy; generated documentation, though improved, may still contain hallucinations or inaccuracies, potentially misleading developers. The underlying LLMs may propagate biases from their training data into the documentation. Over-reliance on such tools could potentially hinder developers' deep code comprehension skills. Applying DocAgent to proprietary code necessitates careful handling, especially regarding external queries, to avoid inadvertently leaking sensitive information. Finally, the computational resources required for LLM-driven multi-agent systems represent a notable cost and environmental consideration. Future work should address these limitations, focusing on robustness, bias mitigation, and deeper evaluation, while emphasizing that human oversight remains crucial in practical deployment.

\bibliography{custom}

\clearpage
\appendix

\section{Related Work}

\textbf{LLM Agent}\ \ Recent advancements in LLM agents have enabled automating complex code-related tasks~\cite{yang2025codereasoning}. Single-agent frameworks like ReAct \cite{yao2022react} and Reflexion \cite{shinn2023reflexion} integrate action-reasoning and self-reflection. Multi-agent systems (CAMEL \cite{li2023camel}, AutoGen \cite{wu2023autogen}) extend these ideas with role-specialized LLMs and structured communication to handle more complex problems. In software development, MapCoder \cite{zhang2023mapcoder}, RGD \cite{chen2023teaching}, and ChatDev \cite{qian2023chatdev} leverage specialized agents for many downstream tasks, achieving state-of-the-art code generation. These insights on multi-agent coordination and workflow structuring underpin our DocAgent framework, which adopts a topologically-aware, tool-integrated multi-agent design.

\textbf{Code Summarization}\ \ Pre-trained encoders such as CodeBERT\cite{feng2020codebert} and GraphCodeBERT~\cite{guo2021graphcodebert} introduced bimodal and structure-aware learning, while encoder-decoder models PLBART~\cite{ahmad2021unified} and CodeT5~\cite{wang2021codet5} unified code generation and summarization. PyMT5~\cite{clement2020pymt5} extended T5 for Python docstring translation with multi-mode support. Recently, LLMs (OpenAI Codex~\cite{chen2021evaluating}, StarCoder~\cite{li2023starcoder}, CodeLlama~\cite{roziere2023code}) have demonstrated strong zero-shot summarization. However, they often lack repository-level context, dependency awareness, and collaboration—limitations our multi-agent, context-aware \textsc{DocAgent} aims to overcome.

\section{Why External Information is needed}
\label{appendix:why_external}

The external open-internet information request is necessary for writing documentation for some novel, newly-proposed ideas, like novel evaluation method, algorithm, model structure, loss functions. For example, DPO~\cite{rafailov2023direct} is a reinforcement learning algorithm proposed in 2023. Codellama has the knowledge cutoff in Sep 2022. So when using codellama for documentation generation, without accessing mathematical intuition and description of DPO from the open internet, codellama will not possible to write helpful documentation that describe the intuition behind the implementation of DPO.

\section{Scarcity of Code Documentation}
\label{appendix:scarcity}

\begin{figure}[ht]
\centering
\begin{minipage}{0.4\textwidth}
  \centering
  \includegraphics[width=\linewidth]{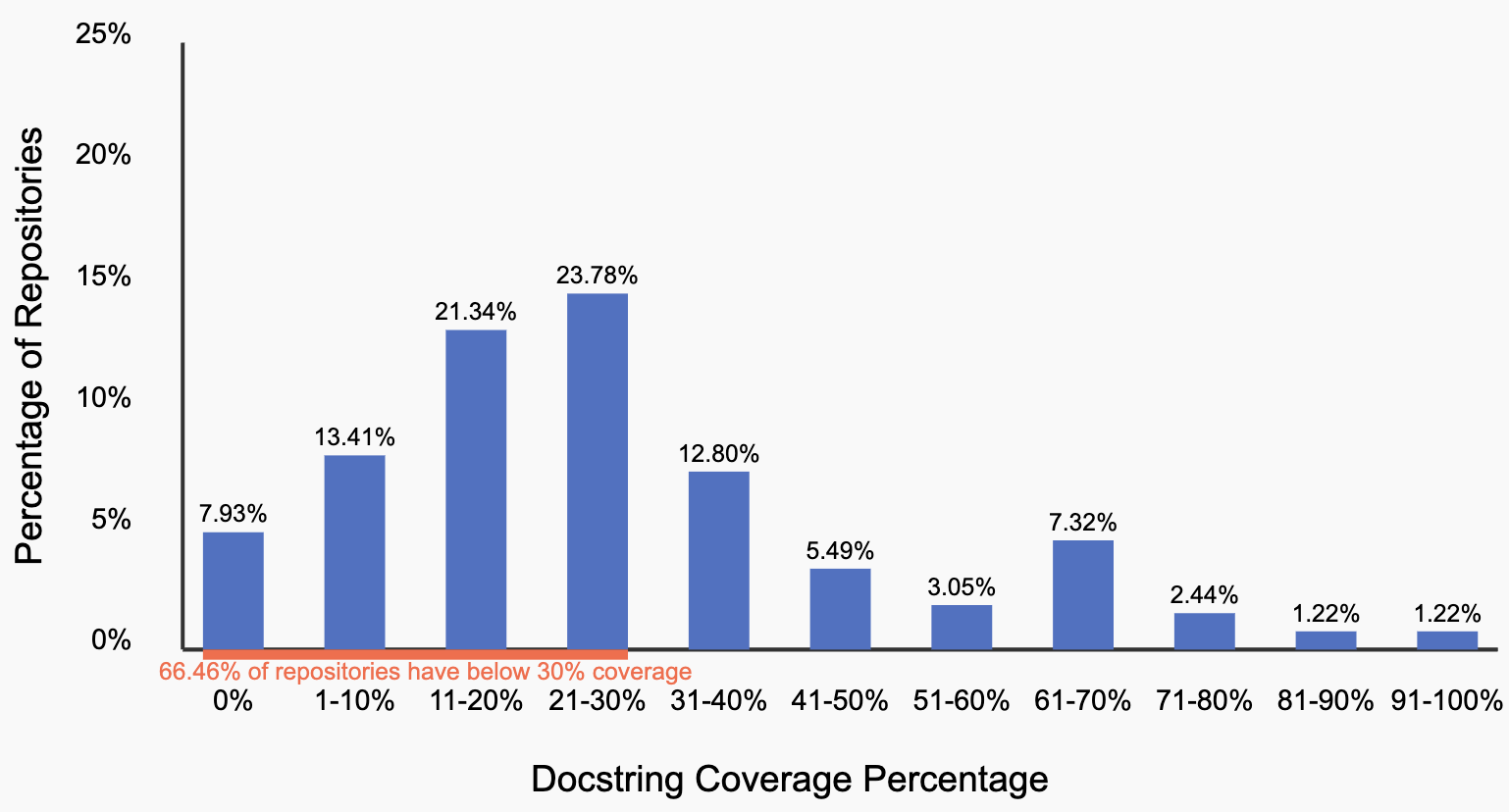}
  \caption{Distribution of repositories by code documentation coverage.}
  \label{fig:coverage}
\end{minipage}
\hfill
\begin{minipage}{0.4\textwidth}
  \centering
  \includegraphics[width=\linewidth]{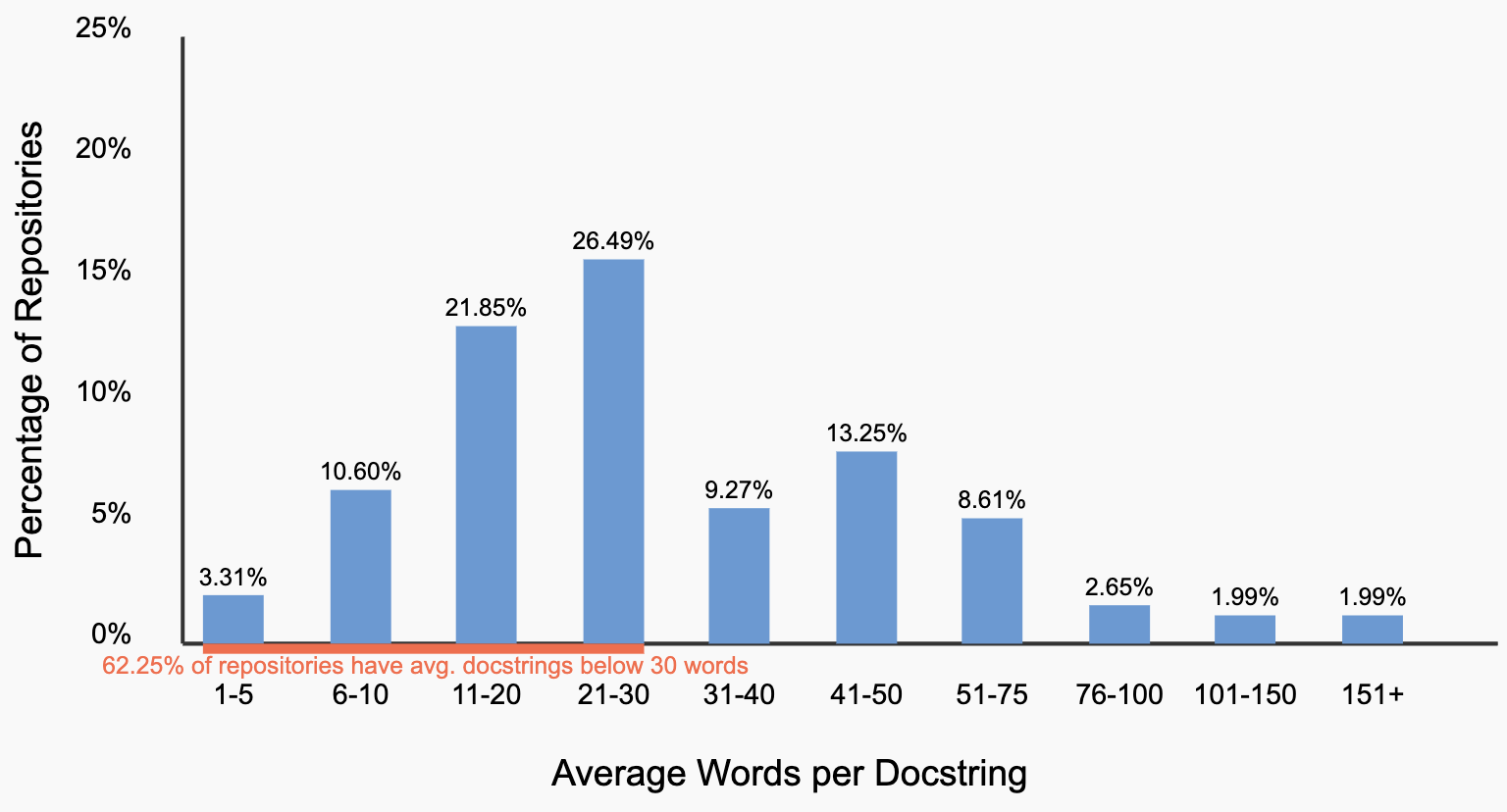}
  \caption{Distribution of repositories by average words per documentation.}
  \label{fig:words}
  
\end{minipage}
 \vspace{-4mm}
\end{figure}

We analyzed 164 top-starred Python repositories (created after January 1, 2025), encompassing 13{,}314 files and 115{,}943 documentable nodes (functions, classes, and methods). Of these nodes, only 27.28\% contained any documentation, with 66.46\% of repositories exhibiting less than 30\% coverage (Figure~\ref{fig:coverage}). Furthermore, 62.25\% of repositories averaged 30 words or fewer per documentation block (Figure~\ref{fig:words}), while only 3.98\% exceeded an average of 100 words, illustrating the widespread brevity and overall scarcity of code documentation.

\section{Data}
\label{appendix:data}
\begin{figure}[h]
    \centering
    \includegraphics[width=0.8\linewidth]{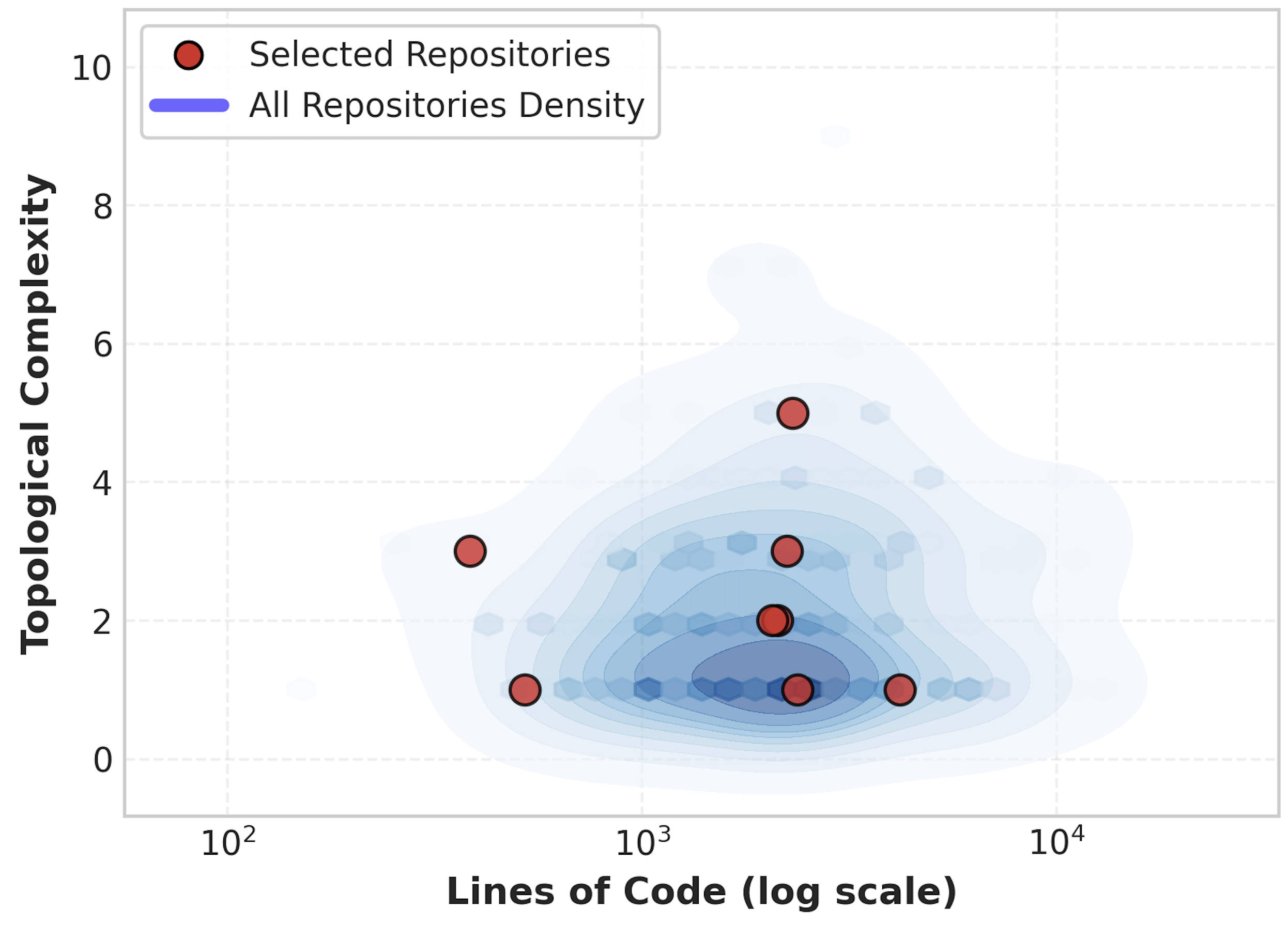}
    \caption{Distribution of repositories used for docstring generation evaluation.}
    \label{fig:data}
    \vspace{-4mm}
\end{figure}

We gathered 164 top-stared Python repositories from GitHub, each created after January 1, 2025, having more than 50 stars, and exceeding 50~KB in size. From this corpus, we selected 9 repositories reflecting the overall distribution in terms of lines of code and topological complexity. Figure~\ref{fig:data} shows the selected repositories (red points) overlaid on the broader distribution. Eventually, we collected 366 code components (120 functions, 178 methods, and 68 classes) for evaluation, with a separate subset of 50 distinct code components (randomly sampled from the full set) used specifically for our truthfulness ablation study.

\section{Robust LLM-as-judge}
\label{appendix:llm_judge_robustness}

Assessing the qualitative aspects of Helpfulness automatically is inherently challenging due to subjectivity. We employ an LLM-as-judge approach, but incorporate rigorous mechanisms inspired by existing work to enhance reliability and consistency, mitigating known issues like positional bias or variability~\cite{wang2024autosurvey, zhuge2024agent}:
\textbf{Decomposed Evaluation:} Instead of a single holistic judgment, the LLM evaluates distinct parts of the docstring (e.g., summary, parameter descriptions, overall description) separately, using tailored prompts for each part~\citep{liu2023xeval, lee2024checkeval}.
\textbf{Structured Prompting:} Each prompt provides the LLM with:
    \begin{itemize}
        \item \textit{Explicit Rubrics:} Detailed criteria defining expectations for different levels on a 5-point Likert scale (1=Poor to 5=Excellent) concerning clarity, detail, and utility specific to the docstring part being evaluated~\citep{kim2023prometheus, zhang2023themis}.
        \item \textit{Illustrative Examples:} Few-shot examples demonstrating good and bad documentation snippets corresponding to different score levels, grounding the rubric criteria~\citep{zheng2023judging, chiang2023alternative}.
        \item \textit{Chain-of-Thought Instructions:} Guiding the LLM to first analyze the code, then compare the corresponding docstring section against the rubric, justify its rating step-by-step, and identify specific strengths or weaknesses~\citep{liu2023geval, zheng2023judging}.
        \item \textit{Standardized Output Format:} Requiring the LLM to output its rating along with detailed justifications in a structured format (e.g., JSON), facilitating aggregation and analysis while ensuring the LLM adheres to the evaluation protocol~\cite{liu2023geval, lee2024checkeval, krumdick2025nofreelabels}.
    \end{itemize}
This structured LLM-as-judge approach aims to provide a scalable yet nuanced assessment of the documentation's practical value to developers.

\section{More System Screenshots}

\begin{figure}[h]
    \centering
    \includegraphics[width=0.9\linewidth]{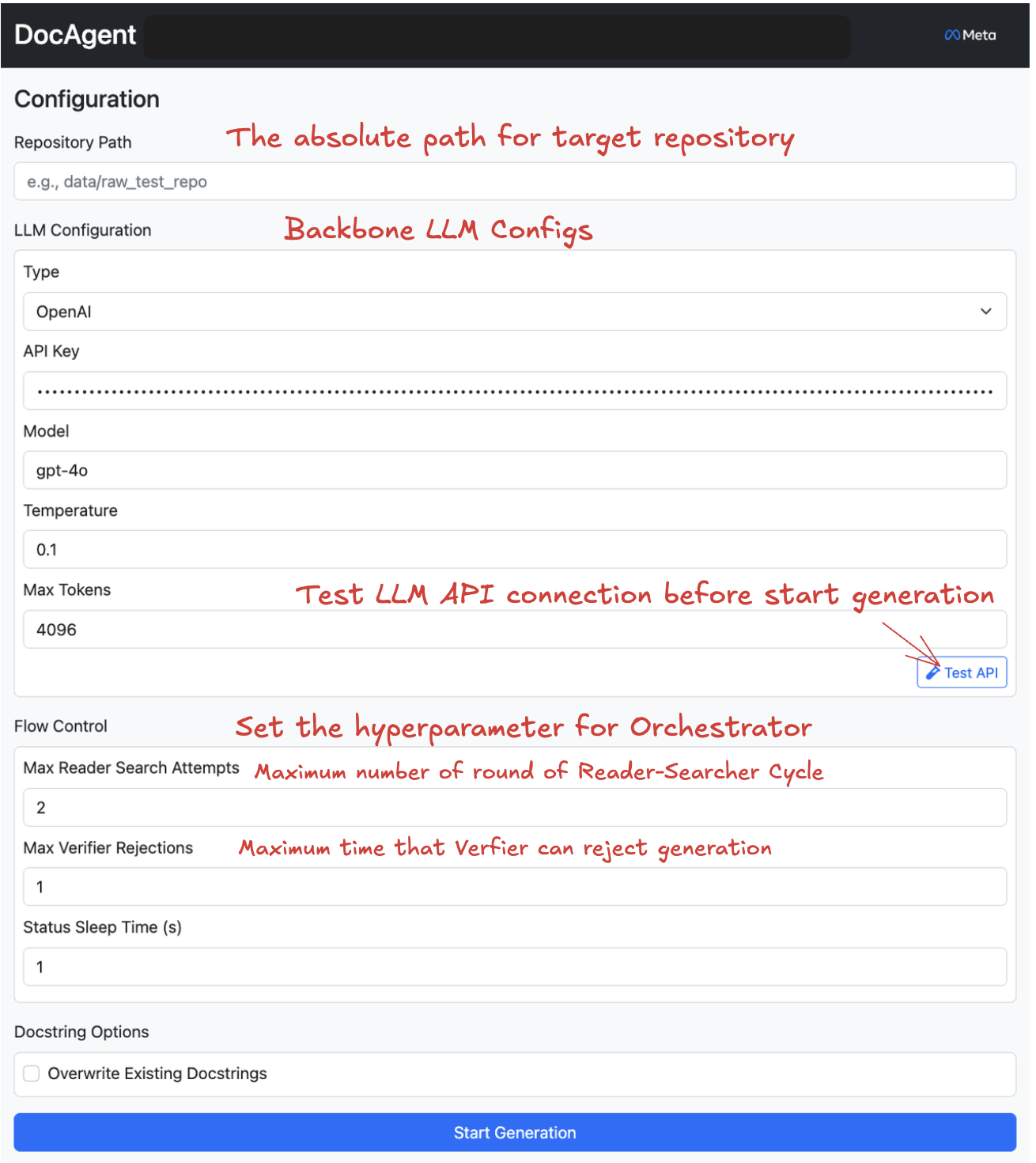}
    \caption{Screenshot of the configuration page.}
    \label{fig:sys_config}
    \vspace{-2mm}
\end{figure}

Figure~\ref{fig:sys_config} shows the configuration page before initiating the code documentation generation process. The page mainly consists of three parts: the target repository path, LLM configuration, and flow control (for the orchestrator).

\begin{figure}[h]
    \centering
    \includegraphics[width=0.99\linewidth]{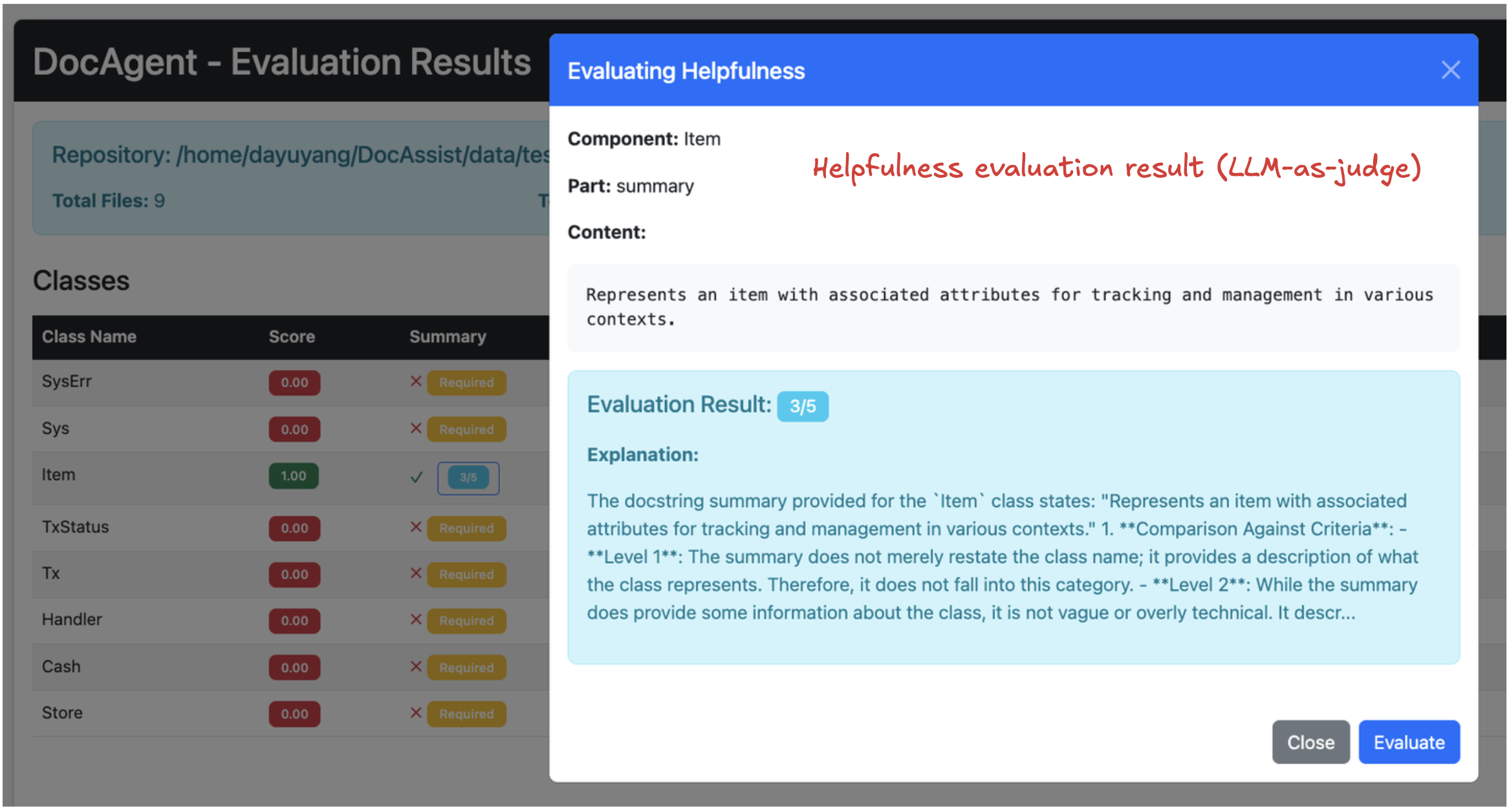}
    \caption{Screenshot of the helpfulness evaluation window.}
    \label{fig:sys_helpfulness}
     \vspace{-4mm}
\end{figure}

Figure~\ref{fig:sys_helpfulness} displays the window that appears after clicking the "Evaluate" button. Since using an LLM as a judge is costly (consuming approximately 500 tokens per evaluation), this feature is optional in the web UI. Only when the user clicks the "Evaluate" button will the evaluation be triggered, after which the button changes to display the generated score.

\end{document}